\documentclass[twocolumn,epsf,psfig]{revtex4-1}
\usepackage{amssymb}
\usepackage{epsfig}
\DeclareMathAlphabet{\pazocal}{OMS}{zplm}{m}{n}   
\usepackage{graphicx}
\usepackage{color}
\usepackage{bm}
\usepackage[normalem]{ulem}    
\usepackage{soul}
\usepackage{amsmath}   
\usepackage{braket}

\begin{document} 
\title{Orbital gyrotropic magneto-electric effect and its strain engineering  in  monolayer  Nb$X_2$ }

\author{Sayantika Bhowal} 
\email{bhowals@missouri.edu}
\affiliation{Department of Physics \& Astronomy, University of Missouri, Columbia, MO 65211, USA}
\author{S. Satpathy}
\affiliation{Department of Physics \& Astronomy, University of Missouri, Columbia, MO 65211, USA}    

\begin{abstract}
Electrical control of the orbital degrees of freedom  is an important area of research in the emerging field of ``orbitronics.'' Orbital {\it gyrotropic} magneto-electric effect (OGME) is the generation of an orbital magnetization in a nonmagnetic 
metal by an applied electric field. 
Here, we show that strain induces a large GME in the monolayer Nb$X_2$ ($X =$ S, Se) normal to the plane, primarily driven by the orbital moments of the Bloch bands as opposed to the 
conventional spin magnetization, without any need for spin-orbit coupling.
 The key physics  is captured within an effective two-band valley-orbital model
 and it is shown to be driven by three key ingredients:  the intrinsic valley orbital moment, broken $C_{3z}$ symmetry, and strain-induced Fermi surface changes.
 The effect   can be furthermore switched by changing the strain condition, with potential for future device applications.

\end{abstract}

\maketitle

Electric field induced magnetization, known as  the magneto-electric effect  \cite{Fiebig}, has been an active area of research for some time.
Conventional magneto-electric effect  involves magnetic insulators and primarily  spin magnetization \cite{LL,Siratori}; however it was recognized that nonmagnetic
metallic systems with certain symmetries
could also show the same effect \cite{Levitov,Edelstein,Kato,Yang}. 
Recently, it was pointed out that ``gyrotropic" (optically active materials with broken inversion ${\mathcal I}$  symmetry) metals having time-reversal ${\mathcal T}$ symmetry show a magneto-electric effect, where in addition to spin, orbital magnetization could also contribute \cite{Souza, Souza_prb, Murakami,Pesin}. The gyrotropic
magneto-electric effect (GME) \cite{note1}
is driven by the intrinsic magnetic moment, spin as well as orbital, of the Bloch electrons at the Fermi surface.
The interest in the GME was revived with the recent advent of the the topological Weyl semimetals, which show a substantial contribution from the orbital magnetization in addition to the spin magnetization \cite{Souza,Souza_prb}. 

 In this paper, we show that by applying strain, a large GME can be induced in the well-known 2D transition metal dichalcogenides (TMDCs) Nb$X_2$ ($X =$ S, Se), originating primarily from orbital
 magnetization with negligible contribution from spin. 
 The TMDCs are excellent for this purpose due to the fact that the
 complex wave functions lead to a robust intrinsic orbital moment at the valley points
 \cite{Xiao2013,ohe2020}, 
 and, while they are 
not gyrotropic (although ${\mathcal I}$ symmetry is broken), they can be engineered to be so with the application of strain, which breaks the $C_{3z}$ symmetry, leading 
 to the orbital GME (OGME). 
 Not only is the OGME strong in Nb$X_2$, but also the direction of the generated orbital magnetization can be reversed
 by changing the strain condition,
 which may have important implication for future  ``orbitronics" device applications.
  Furthermore, the prediction of exotic Ising superconductivity \cite{Xi} in metallic TMDCs offers a new paradigm for superconducting orbitronics.

 \begin{figure}[ht]
\centering
\includegraphics[width=\columnwidth]{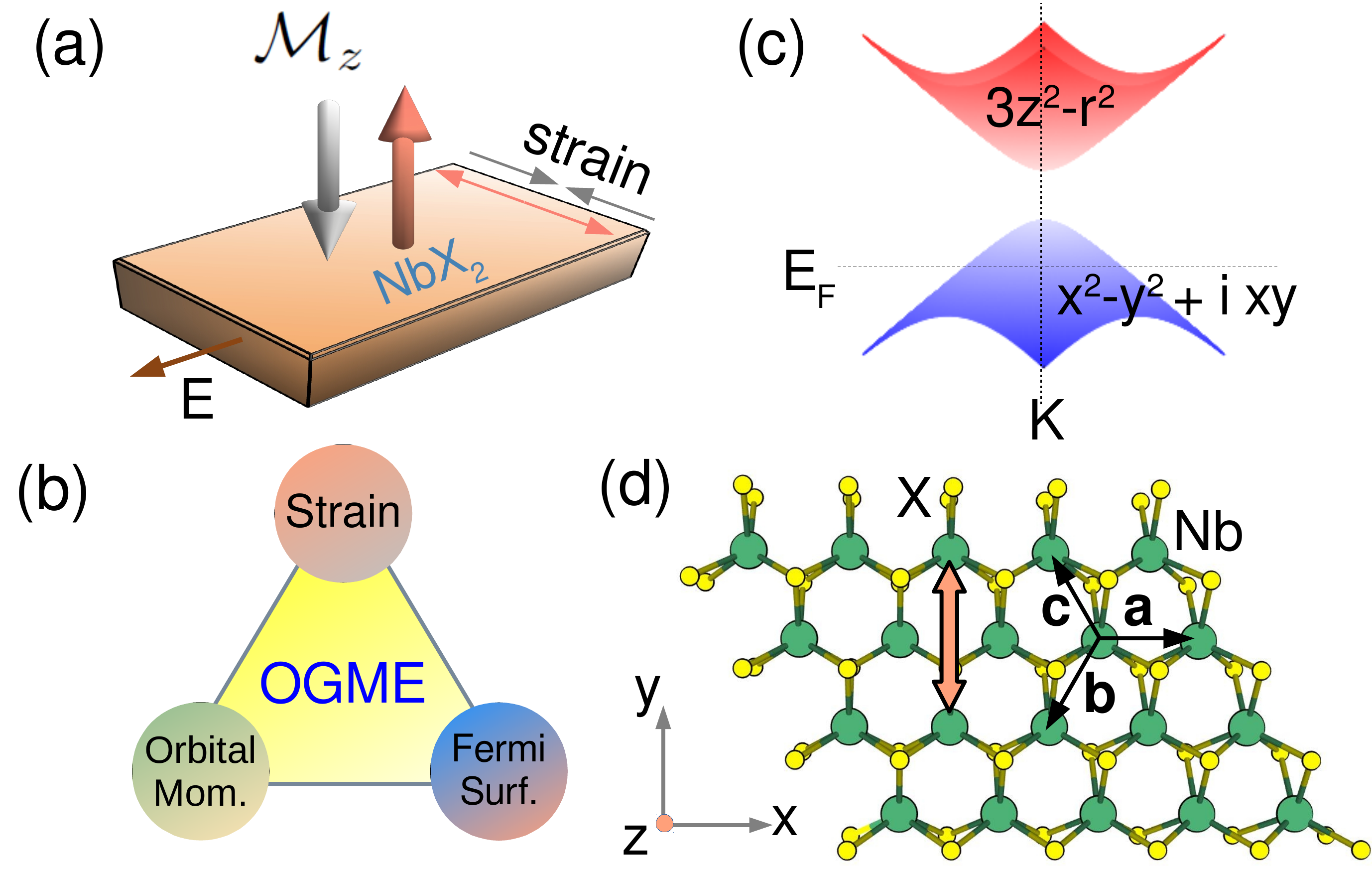}  
\caption { Basic physics of the OGME and its strain manipulation. 
(a) The electric-field induced  orbital magnetization ${\cal M}_z$, which can be tuned and its 
direction reversed by changing the strain.
(b) The three key ingredients for the OGME in the TMDCs, viz.,  uniaxial strain, metallicity, and the existence of an intrinsic orbital moment in the Brillouin zone.
(c) Schematic band structure near a valley point, which contributes the major part to the effect. 
$E_F$ is the Fermi energy.
(d) The 2D crystal structure of NbX$_2$, with the double arrow 
indicating the uniaxial
strain along $\hat y$ considered in the paper. 
}
\label{fig1} 
\end{figure}
%
%

 Monolayer TMDCs have been the subject of intense research due to their valley-spin coupling \cite{Xiao,Ominato}. Several attempts have been made earlier to generate a spin magnetization in monolayer and bilayer TMDCs by creating valley polarization, with an emphasis on spin-orbit coupling (SOC) \cite{PNAS,bilayer}. However, the orbital degrees of freedom in TMDCs are often overlooked. Recently, we have pointed out the crucial role of the valley-orbital locking in generating a large orbital Hall current \cite{ohe2020,ohe_big}. The valley-orbital locking,
 where the two different valleys $K$ and $K'$ have opposite orbital  moments,
  is more fundamental than the valley-spin coupling,
  in the sense that no SOC is necessary for the former.
   The presence of this intrinsic valley orbital moment, although not sufficient for the OGME, provides a fertile ground to observe the effect in TMDCs. 
   The additional ingredient of a gyrotropic symmetry is provided by an applied uniaxial strain, and we find a large OGME, without any involvement of the SOC. The GME due to the 
   spin magnetization is much smaller in comparison.
Since it is a Fermi surface effect, metallic TMDCs, such as Nb$X_2$
constitute an excellent platform for the experimental observation of this effect. 

 The basic physics is illustrated in Fig. \ref{fig1}. Metallic Nb$X_2$   
  crystallizes in a structure with the $D_{3h}$ point group, which is  non-gyrotropic, so that the GME is absent. 
  Explicit breaking of the $C_{3z}$ symmetry makes it gyrotropic with the point group symmetry $C_{2v}$, 
  resulting in a non-zero GME response.
  This is done in the present work by 
  applying a uniaxial strain along the arm-chair direction ($\hat y$). 
  Using valley-orbital model as well as explicit calculations based on  density functional theory (DFT), we show that the strained monolayer TMDCs exhibit a large OGME. For an applied electric field along $\hat x$-direction, an out-of-plane orbital magnetization ${\cal M}_z$ develops, the direction of which can be, furthermore, switched  by changing the strain condition between compressive and tensile. 

 {\it Valley-orbital model--} We consider a minimal two band tight-binding (TB) valley-orbital model, which illustrates the key
 physics of the OGME in the strained TMDCs. DFT studies presented later confirm that the hole pockets at the valley points 
 make the predominant contribution, so that it suffices to focus on the valley points to illustrate the effect.
 By L\"owdin downfolding \cite{downfolding} of the chalcogen orbitals in the TB Hamiltonian and keeping the three important $d$ orbitals for
 the description of the valence and conduction bands, we get the effective Hamiltonian in the presence of strain 
 (see Supplemental Materials \cite{SM} for details). 
 
 The Hamiltonian is
 %
%
\begin{eqnarray} \label{HK}     
{\cal H} (\vec q ) &=& {\cal H}_0 + {\cal S} {\cal H}_1 = \vec d \cdot \vec \sigma + {\cal S}(d_0 \sigma_0 + \vec d_1 \cdot \vec \sigma),
\end{eqnarray} 
where only terms linear  in strain $\cal{S}$ and in momentum
$\vec q = \vec k -\vec K$ or $ \vec k -\vec K'$
have been kept. Here, $\sigma_0$  and $\vec \sigma$ are, respectively, the identity matrix and the Pauli matrices for the pseudo-spin basis
$\ket v = \ket {x^2-y^2} + i\tau \ket{xy}$ and $\ket c = \ket {3z^2-r^2}$, and 
the strain $\cal S$, along the armchair direction,
is by convention positive (negative) for tensile (compressive) strain. 

Note that in Eq. (\ref{HK}), the spin-orbit coupling is omitted, because as shown from the DFT results, discussed later in Fig. \ref{fig4}, the spin contribution to the GME is negligible.
Indeed, as discussed in the Supplementary Materials \cite{SM}, the
SOC term has the Ising form
$ \mathcal{H}_{\rm SOC} =\frac{\tau \lambda} {2}  (\sigma_z+1) \otimes s_z$ in
the two-orbital subspace, $\ket v$ and $\ket c $, and with this form, the spin moments would contribute exactly zero to the GME.
In the DFT
results, the spin contribution is non-zero due to higher
order terms in the Hamiltonian.

The parameters of the Hamiltonian Eq. ({\ref{HK}) are:  
$\vec d = (\tau t q_x a, -t q_y a, -\Delta/2)$, $ d_0 = \tau q_x a \beta_+/2$, and
$ \vec d_1 = (\tau \kappa q_x a + \gamma, \kappa q_y a, \tau q_x a \beta_-/2)$, 
where $a$ is the lattice constant and
$\beta_\pm = \beta_1 \pm \beta_2$.
The valley index $\tau = \pm 1$ for the valley points $K$ and $K'$ respectively.
There are two parameters in the Hamiltonian for the unstrained case, viz., the hopping $t$ and the energy gap $\Delta$, and four additional  parameters that describe the effect of strain, viz., $\beta_1, \beta_2, \gamma, $ and $\kappa$, which are listed in Table \ref{tab1} for NbX$_2$.

 \begin{table} [hb]    
\caption{Hamiltonian parameters (in eV) for  Nb$X_2$, $X =$ S, Se, extracted from DFT calculations.} 
\centering
\setlength{\tabcolsep}{6pt}
 \begin{tabular}{c | c c | c c c c }
 \hline
 Material & \multicolumn{2}{c|}{Unstrained case} &  \multicolumn{4}{c}{Strain parameters} \\ 
 & t & $\Delta$ & $\beta_1$ & $\beta_2$ & $\gamma$ & $\kappa$ \\
 \hline
NbS$_2$  & -0.9 & 1.3 & 3.1 & 0.4  & -2.0 & -0.7 \\ 
NbSe$_2$ & -0.8 & 1.3 & 2.6 & 0.4  & -1.7 & -0.7\\
 \hline     
\end{tabular} 
\label{tab1} 
\end{table}

 The valley-orbital model (\ref{HK}) was derived earlier using a different approach \cite{strain}. 
 Our TB derivation has the advantage that it relates each of the Hamiltonian parameters to some TB hopping integrals, which provide useful insight into the OGME. For example, it is easy to see that
 (Supplementary Materials \cite{SM})  the strain induced parameters $d_0$ and $\vec d_1$ vanish in presence of the $C_{3z}$ symmetry. 
Note that in absence of strain, ${\cal S} = 0$,
the Hamiltonian (\ref{HK}) boils down to  the well known form ${\cal H} (\vec q) =  \vec d \cdot \vec \sigma $, representing the massive Dirac particle.

The Hamiltonian  (\ref{HK}) allows for an analytical calculation of the orbital magnetic moment and the OGME.
Diagonalization yields the energy eigenvalues 
\begin{equation}\label{energy}
 \varepsilon_{\pm} (\vec q) = \frac{1}{2} [\tau {\cal S}\beta_{+} q_x a \pm \{ (\Delta -\tau {\cal S}\beta_{-} q_xa)^2+ 4 d^2 \}^{1/2}],
 \end{equation}
where $\pm$ correspond to conduction and valence bands respectively, $d^2 = (d_x^{2}+d_y^{2})$,
$d_x =\tau t_x q_xa + \gamma {\cal S} $, 
 $d_y =-t_y q_ya$, and $t_{x,y} = (t \pm \kappa {\cal S})$. 
It directly follows from Eq. (\ref{energy}) that the constant energy ($\varepsilon$) contours of the valence band are elliptical in shape, viz., 
 \begin{equation}\label{ellipse}
    \frac{(q_x+\tau q_s)^2}{\gamma_x^2} + \frac{q_y^2}{\gamma_y^2} =1,
  \end{equation}
with the center of the ellipse  shifted by the amount $\tau q_s$ from the valley points along $\hat x$ (perpendicular to the  strain direction $\hat y$), where
$q_s =
\lambda (\varepsilon) {\cal S} + O({\cal S}^2)$, with $\lambda (\varepsilon) = (4t^2 a)^{-1}  \times ( 2 \varepsilon \beta_+ - \Delta\beta_-+4t \gamma) $.
The shift, which is opposite for the two valleys and also for the two strain conditions (tensile vs. compressive), is important for the net OGME and
plays a key role in the strain switching. 
The ratio of the two semi-axes is $\gamma_y/ \gamma_x = 1  + 2 \kappa t^{-1} {\cal S} $, so that the
 ellipse is elongated along $\hat x$ or $\hat y$ depending on the sign of ${\cal S}$.
 The elliptical shapes are confirmed from the DFT results, presented later in Fig. \ref{fig3}.

The magnetic moment  $\vec M (\vec k)$ for the valence band, needed for the OGME response function, 
are computed from the band energies $\varepsilon (\vec k) $ and the wave functions $u (\vec k) $
using the expression  \cite{Souza,Murakami}
  \begin{eqnarray} \label{mom} \nonumber
  \vec M (\vec k) 
  &=&  \frac{e}{2\hbar} {\rm Im} \bra {\vec \nabla_k u(\vec k)} \times [{\cal H}(\vec k) -\varepsilon (\vec k)] \ket{\vec \nabla_k u(\vec k)} \\  \nonumber 
  &+&  \frac{e}{\hbar} {\rm Im} \bra {\vec \nabla_k u(\vec k)} \times [\varepsilon (\vec k) -E_F]  \ket{\vec \nabla_k u(\vec k)}  \\ 
 &-&  \bra{u(\vec k)} {\vec m_s} \ket{u(\vec k)},
 \end{eqnarray}
  where the first two terms are the orbital moment contributions, computed by evaluating the expectation value of the orbital magnetization operator  $(-e/2) \vec r \times \vec v$, \cite{Niu} where $-e < 0$ is the electronic charge. The first term in Eq. (\ref{mom}) is due to the self-rotation, and the second is due to the center-of-mass motion of the wave packet \cite{Niu},  while the third term is the spin moment, with  
 $\vec m_s =  g_s \mu_B \vec s $   
   being the spin moment operator.
However, as argued earlier, we omit the spin contribution
as it hardly contributes to the GME.
Furthermore, since 
the OGME response function is a Fermi surface property, the second term in Eq. (\ref{mom}) does not contribute also, so that we proceed to compute the first term in our model. 
 %
  %
  %
    %
 %
 
  %
  %
  %
 %

 \begin{figure}[tb]
\centering
\includegraphics[scale =0.33 ]{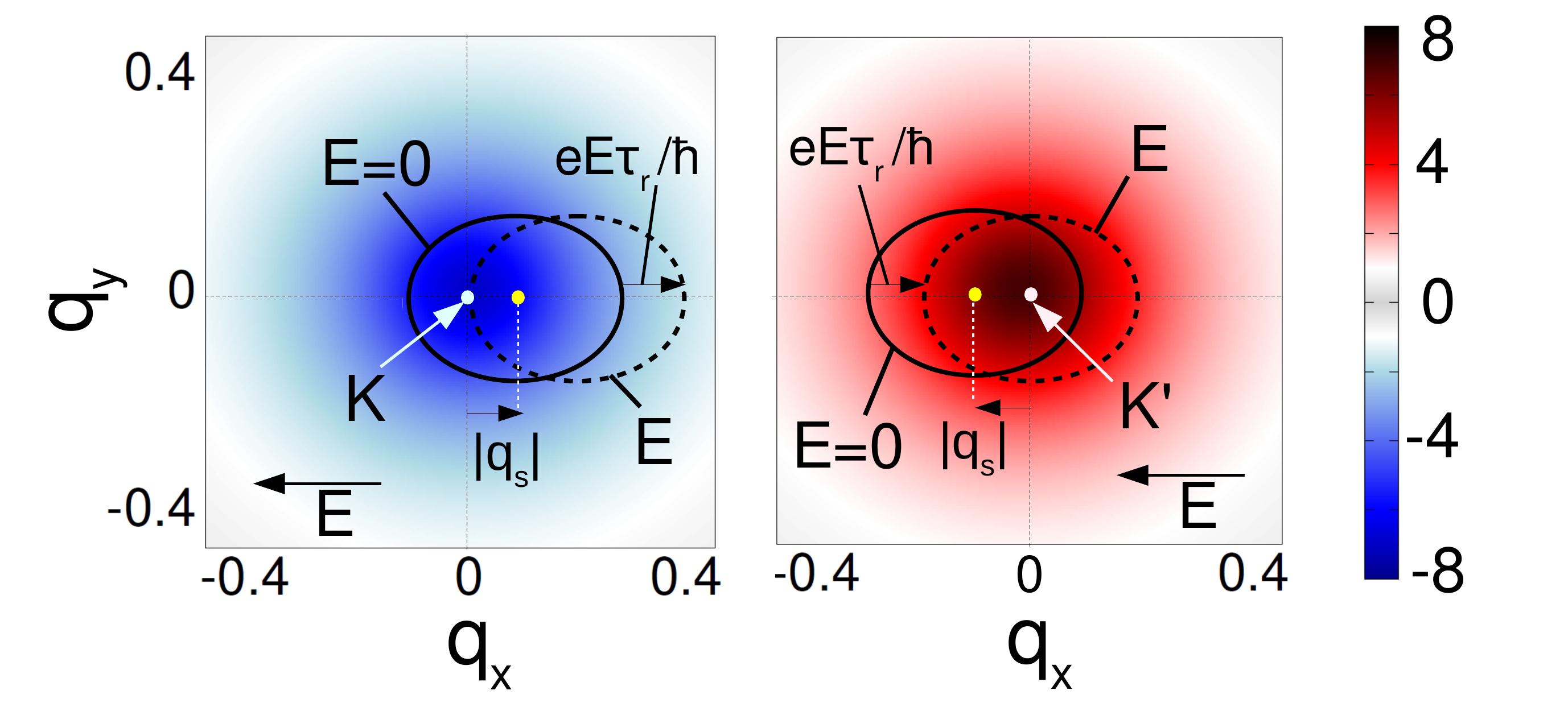}  
\caption { Illustration of the orbital moments (color gradation) near the valley points $K$ and $K'$ with compressive strain, 
obtained from Eq. (\ref{mz}), and the elliptic Fermi surface (solid ellipse) with its center shifted by  $\tau q_s$, 
Eq. (\ref{ellipse}), from the valley points. 
The electric field displaces the Fermi surface by $e E \tau_r/\hbar$ (dashed ellipse), leading to a 
net orbital magnetization  and the OGME.
The Fermi surface for the unstrained case is circular (not shown), as may be inferred from Eq. (\ref{ellipse}). 
}
\label{fig2} 
\end{figure}

The only non-zero component of the orbital moment, $M_z (\vec q)$, may be computed analytically from 
Eq. (\ref{mom}), 
using the eigenvalues Eq. (\ref{energy}) and the corresponding wave functions. 
After a  tedious but straightforward calculation, we find the result
for the valence states near valley points
\begin{eqnarray}\label{mz} \nonumber
M_z(\vec q) &=& -\frac{\tau  t_ya^2 (t_x\Delta  + \beta_- \gamma {\cal S}^2) }{(\Delta-\tau{\cal S}\beta_-q_xa)^2+4(d_x^2+d_y^2)} \\ 
&=&M^0_z (\vec q)+ A (\vec q) {\cal S} + O({\cal S}^2) + ...,
%
%
 \end{eqnarray}
where  $M^0_z  (\vec q)= -\tau t^2a^2 \Delta (\Delta^2 + 4t^2q^2a^2)^{-1}$ is the moment without the strain, and  
the linear strain coefficient 
$A (\vec q)  = f(q^2) \times [\eta q_x  + \tau \eta' (q_x^2-q_y^2)]$, where $\eta = 2a (4\gamma t - \Delta \beta_-)$, $\eta' = 8t\kappa a^2$, and $f(q^2) = t^2a^2 \Delta (\Delta^2 + 4t^2q^2a^2)^{-2}$.
%
%
Note that $M_z$ has opposite signs at the two valleys ($\tau = \pm 1$), as seen from Fig. \ref{fig2},  which comes from the factor $\tau$ in Eq. (\ref{mz}).
As a result of this, the total orbital magnetization is zero both in presence or absence of strain.
When {\it both} an electric field and strain are present, strain-induced shift of the elliptical hole pockets results in an asymmetric distribution of $M_z$ at the two valleys with a net orbital magnetization,
proportional to $E$, leading to a linear response that we now proceed to calculate.


 The magnitude of the OGME response function $K_{ij}$,
 defined as the orbital magnetization induced by unit electric field, 
 $ {\cal M}_j = K_{ij} E_i$,
 where subscripts denote the cartesian directions,
  may be obtained from the 
  change in the net orbital magnetization with the electric field
 \begin{eqnarray} \label{DM}     
 \vec {\cal M} 
= \frac{1}{(2\pi)^2} \int_{\rm BZ}  d^2 k    \vec M (\vec k) \times
 [ f (\varepsilon_{\vec k + a_0 E \hat k_i}) - f (\varepsilon_{\vec k})], 
  \end{eqnarray}
  where $f$ is the Fermi function, and the relaxation-time approximation has been used
for the non-equilibrium electron distribution, with the Fermi surface shifted by the 
amount  $a_0 = -e \tau_r E / \hbar$ in the direction of the electric field. Here, $\tau_r$ is the relaxation time, and the integral is over the Brillouin zone (BZ). 
A  Taylor series expansion of (\ref{DM}) yields a convenient form for the OGME linear response $K_{ij}$,
which will be employed in the DFT calculations.
The scaled response  
$  {\tilde K}_{ij} \equiv  (\hbar / e\tau_r)  K_{ij} $ reads
  %
  \begin{eqnarray} \label{kme2}
 {\tilde K}_{ij} &= & \frac{1}{(2\pi)^2} \int_{\rm BZ} d^2 k  \    v_i (\vec k) \ M_j (\vec k)\Big(-\frac{\partial f_0}{\partial \varepsilon} \Big)_{\varepsilon(\vec k)},
 \end{eqnarray}
where 
$ \vec v (\vec k) =  \vec \nabla_k  \varepsilon (\vec k) $ is the electron velocity.

The response function ${\tilde K}_{ij}$ can be calculated by computing the change in orbital magnetization due to the shifted Fermi surfaces in Fig. \ref{fig2}  using Eq. (\ref{DM}), or from Eq. (\ref{kme2}). 
We take $T = 0$. Since the effect is absent for a completely filled band, OGME for the electrons is equal and opposite to that of the holes. Note that, since the elliptical hole pockets are shifted along $\hat  x$ due to strain, electric field along $\hat y$ does not change the net orbital moment due to the cancellation from the two valleys,  so that   ${\tilde K}_{yz} = 0$
and ${\tilde K}_{xz}$ is the only remaining non-zero component.
To compute this using  Eq. (\ref{DM}), it is convenient to first write the orbital moment (\ref{mz}) in terms of momentum $\vec q$ measured with respect to the ellipse center and then integrate over the four crescents in Fig. \ref{fig2},
produced by the displacement of the Fermi surface by the electric field.
The calculation is straightforward and 
the result is 
\begin{eqnarray} \label{ana}
%
\tilde K_{xz} =  {\pi}^{-1} {\cal S} q_F^2 f(q_F^2)[8t^2a^2 \lambda (E_F)- \eta ] + O ({\cal S}^2),
\end{eqnarray}
where $E_F$ is the Fermi energy and $q_F$ is the Fermi momentum.
The spin degeneracy of the bands and the contribution from both valleys are already included in this expression.

The model response function Eq. (\ref{ana})  contains the central points of our work, viz., first, that the OGME response is non-zero only in the presence of strain, and second, that the response  switches its sign between the tensile and compressive strains,
so that the 
magnetization can be reversed by changing the strain condition.
For an insulating system ($q_F = 0$), we recover the expected result that $\tilde K_{xz} = 0$. 
Furthermore,   
the OGME response increases with $q_F$, i. e., with increasing number of holes, in a wide range  ($q_F \lesssim  (2 t a)^{-1} \Delta $).
 This provides an additional means of tuning by controlling the hole concentration near the valley points, which may also be relevant for the insulating TMDCs with hole doping.

\begin{figure}[t]
\centering
\includegraphics[width=\columnwidth]{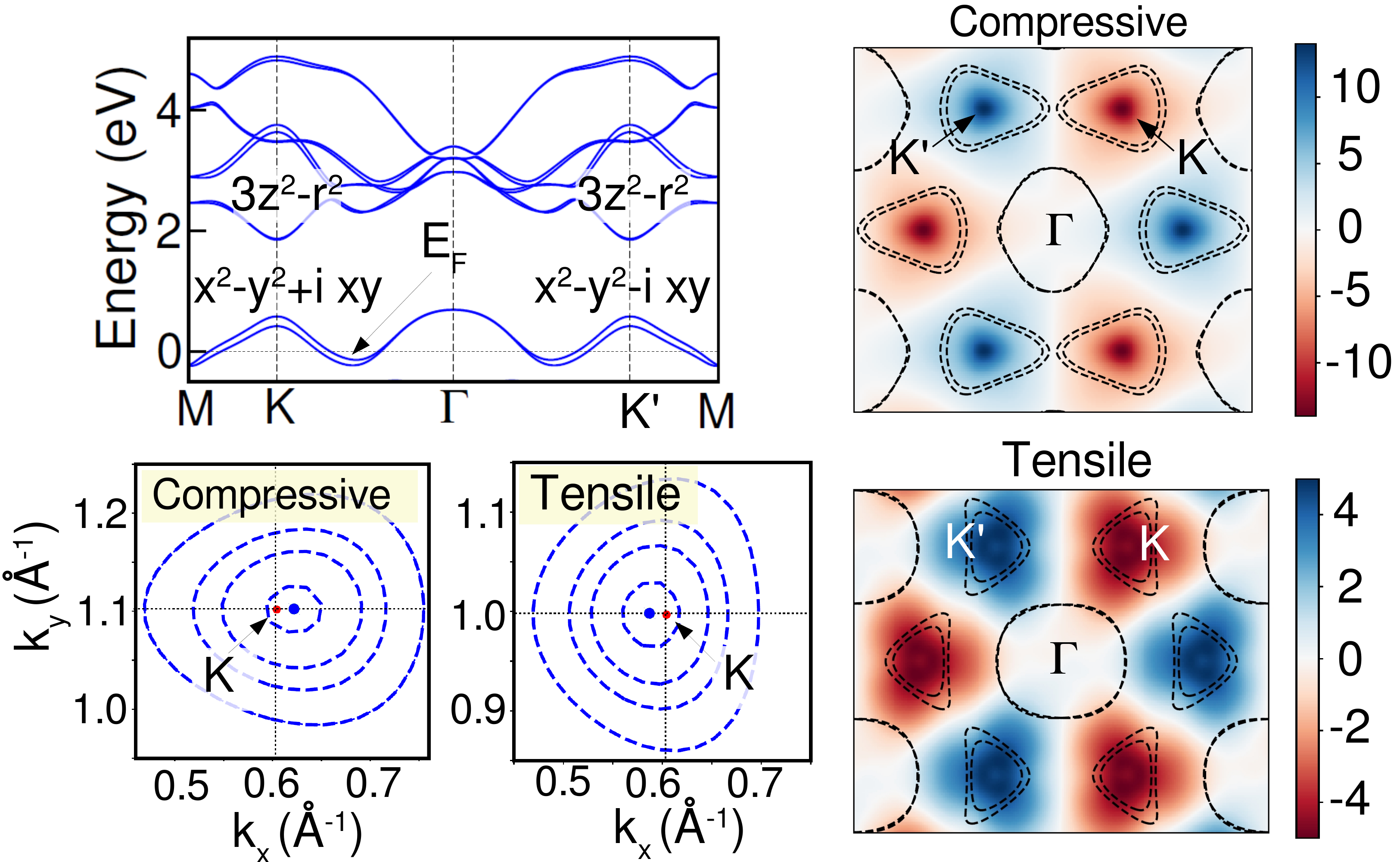} 
\caption { DFT results for NbX$_2$.
{\it Top left,}  
Band structure for the unstrained NbSe$_2$, with circular energy contours (not shown) around the $K$ and $K'$ valence band maxima. The small spin-orbit splitting of the valence bands near the valley points 
is also visible.
{\it Bottom left,}   Elliptical energy contours for the strained case, with the ellipse centers shifted  by strain.  
{\it Right,} Orbital moment for the valence bands in strained NbS$_2$,
computed from DFT using Eq. (\ref{mom}) and summing over all occupied bands.   
The dashed contours indicate the hole Fermi surfaces (spin-orbit split). 
In Eq.  (\ref{mom}), $E_F$ was taken as the valence band  maximum, and 
${\cal S} = \pm 5\%$ was used. 
}
\label{fig3} 
\end{figure}
%

  {\it Density-functional  results--}  
  The basic points of the results of the model are validated from the DFT calculations, which we present below.
Orbital moments and the response function ${\tilde K}_{xz}$  were computed using  Eq. (\ref{kme2})
 for monolayer Nb$X_2$, both in absence and presence of strain, using the first-principles pseudopotential method \cite{QE} and the Wannier functions as implemented in the WANNIER90 code \cite{MLWF,w90}. 
Technical details 
are given in the Supplemental Materials \cite{SM}. 

Fig. \ref{fig3} shows the metallic Nb-$d$ bands in Nb$X_2$, with the half filled valence band,
the complex orbital characters, and the elliptical energy contours, with centers shifted from the valley points in the presence of the strain. The figure also shows that the valley hole pockets, at $K$ and $K'$, have the predominant contribution to the orbital moments, with very little contribution from the $\Gamma$ pocket. 
This validates the valley orbital model Eq. (\ref{HK}), which was developed for the $K$, $K'$ points only.

The metallic TMDCs considered here are non-magnetic, both with and without strain, and this is confirmed by the DFT calculations as well. This being the case, even though the orbital moments are present at individual momentum points in the BZ, they add up to zero, simply due to the time-reversal symmetry. 
In the unstrained case, an applied electric field also fails to develop a magnetic moment, so that the OGME is zero.  
Physically, the vanishing of ${\tilde K}_{ij}$ can be understood by
considering  
 three momentum points at a time on the Fermi surface around a particular valley, which are related to each other by $C_{3z}$ symmetry of the structure. For these three $k$ points,  $M_z$ is the same,
  but the Fermi surface velocities add up to zero,
$\sum_{\alpha =1}^3 \vec v (\vec k_\alpha) = 0$, so that their net contribution to the response expression Eq. (\ref{kme2}) cancels out leading to ${\tilde K}_{ij} = 0$.

The uniaxial strain results in a structural transition from $D_{3h}$ to the gyrotropic point group $C_{2v}$, the $C_{3z}$ symmetry is broken,
 and the system acquires a non-zero OGME, ${\tilde K}_{xz} \ne 0$ as a result.
However,
${\tilde K}_{yz}$ continues to be zero due to the $C_{2y}$ symmetry, still present in the strained structure, 
under which  
the velocities and the moments  transform as
$(v_x, v_y) \rightarrow  (-v_x, v_y)$ and $M_z \rightarrow  -M_z$,
leading to the cancellation of the contributions from the different
valleys, as may be inferred from Eq. (\ref{kme2})

\begin{figure}[t]
\centering
\includegraphics [scale =0.30] {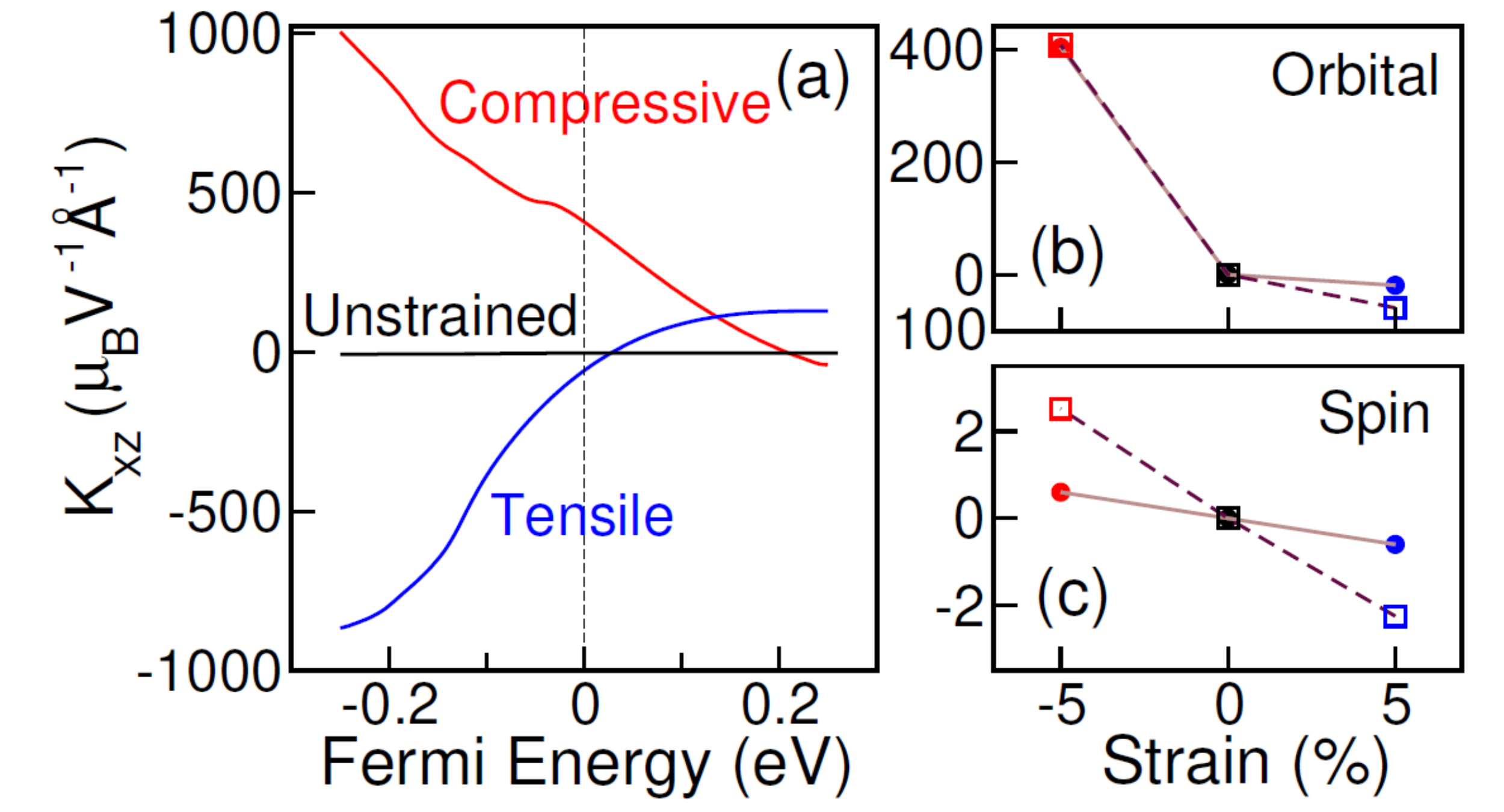}
\caption { (a) The OGME response function $K_{xz}$ for NbSe$_2$ as a function of the Fermi energy. $E_F = 0$ for the undoped NbSe$_2$. (b) and (c)  Orbital and spin contributions to the GME, clearly showing the dominance of the former by two orders of magnitude. The solid circles and the open squares correspond to NbS$_2$ and NbSe$_2$ respectively.     
}
\label{fig4}  
\end{figure}

The response function $K_{xz}$, computed from  DFT as a function of $E_F$,  is shown in Fig. \ref{fig4}, where we have taken a typical relaxation time  $\tau_r \approx 10$ ps \cite{Brida}. The DFT results make several points. One, 
 the OGME is non-zero only when the strain is present, and it is opposite in sign for compressive vs. tensile strain as
predicted from the valley-orbital model result, Eq. (\ref{ana}). Two, the GME arising out of the spin moments is two orders of magnitude smaller than the same arising out of the orbital moments, providing  justification to the neglect of the spin moment in the valley-orbital model. Third, the magnitude of  $K_{xz}$  increases with the  number of holes (larger $E_F$ measured from the valence band top) as predicted from Eq. (\ref{ana}). 
Note that for a completely filled band, $K_{xz} = 0$, which is not shown in Fig. \ref{fig4}.

The magnitudes of $K_{xz}$, computed from the DFT, are of the same order as the model results.  
Using Eq. (\ref{ana}) with parameters in Table \ref{tab1} and $q_F$ corresponding to the hole concentration  of $n_h \approx 0.25$ in each valley,
for NbSe$_2$ we get 
$K_{xz} = \pm 125~ \mu_B$V$^{-1}$\AA$^{-1}$ for compressive and tensile strains, respectively, while the  DFT results are asymmetric for the two strains (407 vs.  -59 ) (as seen from Fig. \ref{fig4} for $E_F = 0$). These differences may be attributed to the anisotropic warping \cite{warping} of the DFT bands as one goes away from the valley points  (see energy contours of Fig. \ref{fig3}, {\it right})
 as well as  to the change in the number of valley holes with strain due to the relative shift
 between the $\Gamma$ and the valley point energies (see Supplementary Materials \cite{SM}). These effects were not included in the model for simplicity. The valley-orbital model, nevertheless, captures the essential physics of the OGME.

The orbital magnetization for NbSe$_2$, computed from the DFT response function, for the case of 
${\cal S} = 5\%$   and $ E = 10^4$ V/m,  is ${\cal M}_z \approx  0.04~ \mu_B/nm^2$. 
We note that this is at least one order of magnitude larger than the reported spin magnetization in the strong Rashba  systems \cite{ingrid}
or the  Weyl semimtals \cite{wsm}, 
 and it is also twice as large as the recently reported magnetization in twisted bilayer graphene \cite{TBG}. Not only is the orbital magnetization  large in NbX$_2$, but it is also switchable by strain, as we have shown.

 To summarize, we have shown that a switchable large OGME can be induced by strain in the metallic TMDCs. 
 A simple valley-orbital model was developed to capture the essential physics of the OGME.  
  The broken $C_{3z}$ symmetry due to the strain leads to the shifting of the hole pockets in the momentum space, 
which both describes the generation of the effect as well as its strain switching. 
 The orbital magnetization may be probed experimentally from magneto-optical Kerr measurements \cite{MOKE} by growing the material using flexible substrates \cite{Flexiblestrain}, or, alternatively, using piezoelectric substrates which may also be optimal for practical devices.
Our work should stimulate search for materials with large OGME for fundamental science as well as for future ``orbitronics" applications. 
 

%


{\it Acknowledgment-- } We thank the U.S. Department of Energy, Office of Basic Energy Sciences, Division of Materials Sciences and Engineering for financial support under Grant No. DEFG02-00ER45818.

\begin{widetext}

\newpage

\begin{center}
{\bf\large Supplementary Materials for \\
Orbital gyrotropic magneto-electric effect and its strain engineering in  monolayer  Nb$X_2$ }
\end{center}

\section{Derivation of the valley-orbital model in presence of strain} \label{TB}

In this section, we discuss the derivation of the valley-orbital model Hamiltonian [Eq. (1) of the main text] for the monolayer transition metal dichalcogenides (TMDCs) in the presence of uniaxial strain, starting from the  tight binding (TB) model for $d$ orbitals. The model Hamiltonian is valid near the valley points $K$ and $K'$, which for all practical purposes control 
the electronic properties of the TMDCs.
We focus here on uniaxial strain along the arm-chair direction ($\hat y$ direction in Fig. \ref{fig1}). The ideas can be easily extended for strain in any general direction and the TB parameters shown in Fig. \ref{fig1} ({\it middle}) are all that will be needed.

To build the model, we retain just the three orbitals (viz., $xy, 3z^2-r^2, x^2-y^2$) on the metal atoms, which describe the valence and the conduction bands around the two valley points. Only metal atoms (M) are kept in the model. However, to include the effect of the broken inversion ($\cal I$) symmetry, we must 
fold in the effect of the ligand atoms (X = S or Se) into the effective M-M hopping integrals. This can be done in two ways: (i) By L\"owdin downfolding \cite{downfolding} of the standard real-space TB hopping integrals \cite{Harrison}, which we have described explicitly for the TMDC's in our earlier work \cite{ohe_big} and (ii) By directly taking the M-M hopping integrals from band structure codes such as the NMTO code \cite{nmto}, where the L\"owdin downfolding is implemented in the momentum space and the relevant bands are fitted to produce the effective hopping parameters. 
Either approach results in the same model Hamiltonian, since the effective parameters are anyway fitted to the band structure near the valley point. Here, we will use the second approach and just take the parameters produced by the NMTO code.

The TB Hamiltonian is written in the Bloch function basis 
%
$
c^\dagger_{\vec{k} m} = \frac{1}{\sqrt{N}} \sum_{i} e^{i \vec{k}\cdot \vec{R}_i} c^\dagger_{im},
$
where $\vec{k}$ is the Bloch momentum and $c^\dagger_{im}$ creates an electron at the
$i$-th site in the orbital $m $, written in
 the order: $xy, \ 3z^2-r^2, $ and $x^2-y^2$. The Hamiltonian reads
\begin{eqnarray}     \label{H}  
{\cal H} (\vec k ) &= 
\left[
{\begin{array}{*{20}c}
    h_{11} & h_{12} & h_{13}   \\
    h^*_{12} & h_{22} & h_{23}  \\
    h^*_{13} & h^*_{23} & h_{33} \\
\end{array} }  \right],
\end{eqnarray} 
where 
\begin{eqnarray} \nonumber
 h_{11} &=& \varepsilon_1 + 2t^a_1\cos k_a + 2t^b_1(\cos k_b + \cos k_c)\\ \nonumber
 h_{12} &=& 2it^a_2 \sin k_a + i(t^b_2+t^c_2) (\sin k_b+\sin k_c) 
 + (t^b_2-t^c_2) (\cos k_b - \cos k_c) \\ \nonumber
  h_{13}&=&  2it^a_3 \sin k_a + i(t^b_3+t^c_3) (\sin k_b+\sin k_c) \
 + (t^b_3-t^c_3) (\cos k_b - \cos k_c) \\ \nonumber
 h_{22} &=& 2t^a_4 \cos k_a + 2t^b_4(\cos k_b + \cos k_c) \\\nonumber
 h_{23}&=&  2t^a_5 \cos k_a + (t^b_5+t^c_5) (\cos k_b + \cos k_c) 
 + 2i(t^b_5-t^c_5) (\sin k_b-\sin k_c) \\ 
 h_{33} &=& \varepsilon_2 + 2t^a_6 \cos k_a + 2t^b_6(\cos k_b + \cos k_c).
\end{eqnarray}
Here, $k_\eta = \vec k \cdot \vec \eta$ and  $\vec \eta = \vec a, \vec b, \vec c$ denote the directions of the neighboring Nb atoms with respect to the Nb atom at the centre, as illustrated in Fig. \ref{fig1}, i.e., $ \vec a = a \hat x,~ \vec b = -(a \hat x/2 + f a \hat y)$ and $ \vec c = -a \hat x/2 + fa \hat y$, $a$ is the Nb-Nb distance along $\hat x$, while the factor $f$ depends on the strain and has the values $f = \sqrt{3}(1 + {\cal S})/2$, where $\cal S$ is the strain along the arm-chair direction $\hat y$, as shown in Fig. 1, {\it left}, and it takes $\pm$ values corresponding to tensile and compressive strain respectively.

We then expand the Hamiltonian (\ref{H}), around the valley points $K$ (-4$\pi$/3a, 0) and $K'$ (4$\pi$/3a, 0) keeping only terms that are linear in $q$, 
where $q \equiv k -K$ or $k-K'$. The resulting Hamiltonian is
\begin{eqnarray}     \label{3band}  
{\cal H} (\vec q ) &= 
\left[
{\begin{array}{*{20}c}
    h^q_{11} & h^q_{12} & h^q_{13}  \\
    (h^q_{12})^* & h^q_{22} & h^q_{23}  \\
    (h^q_{13})^* & (h^q_{23})^* & h^q_{33} \\
  \end{array} }  \right],
\end{eqnarray} 
where

\begin{eqnarray} \nonumber
 h^q_{11} &=& \varepsilon_1-(t^a_1+2t^b_1)+ \tau \sqrt{3}q_xa (t^b_1- t^a_1)\\ \nonumber
  h^q_{12} &=& \tau i \sqrt{3} (t^a_2 +t^b_2+t^c_2) -iq_xa\{ t^a_2 -2^{-1}(t^b_2 + t^c_2) \} + \sqrt{3} \tau f q_ya (t^b_2-t^c_2)  \\ \nonumber
  h^q_{13}&=& \tau i \sqrt{3} (t^a_3 +t^b_3 +t^c_3) - iq_xa\{ t^a_3 -2^{-1}(t^b_3 + t^c_3) \} + \sqrt{3} \tau f q_ya (t^b_3-t^c_3) \}  \\ \nonumber
   h^q_{22} &=& -(t^a_4+2t^b_4)+ \tau \sqrt{3}q_xa (t^b_4- t^a_4)\\ \nonumber
 h^q_{23}&=&  -(t^a_5 +t^b_5+t^c_5) + \frac{\tau \sqrt{3} q_xa}{2} (-2t^a_5 +t^b_5+t^c_5) + ifq_ya (t^b_5-t^c_5) \\ \nonumber
 h^q_{33} &=& \varepsilon_2-(t^a_6+2t^b_6)+\tau \sqrt{3}q_xa (t^b_6- t^a_6),
\end{eqnarray}
and $\tau = \pm 1$ is the valley index
for the $K$ and $K^\prime$ valleys respectively. 
Now, it is well known that in the TMDCs, the valence and conduction bands near the valley points are well described by the wave function characters, $\ket v = (\sqrt{2})^{-1} (|x^2-y^2 \rangle + i \tau |{xy} \rangle)$ and $\ket c = |3z^2-r^2 \rangle$. 
Within this subspace,   Hamiltonian (\ref{3band}) 
becomes a $2 \times 2$ matrix
 \begin{eqnarray}    \label{HK}  
{\cal H} (\vec q ) = 
\left[
{\begin{array}{*{20}c}
    -\frac{\Delta}{2} + \tau \alpha_1 q_xa & \tau t_x q_xa + it_yq_ya + \delta  \\
     \tau t_x q_xa - it_yq_ya + \delta &  \frac{\Delta}{2} + \tau \alpha_2 q_xa\\
   \end{array} }  \right],
   \end{eqnarray} 
where the 
parameters of the Hamiltonian (\ref{HK})
 can be expressed in terms of the TB hopping parameters. The relations are:
\begin{eqnarray}    \nonumber     \label{para} 
 t_x &=& (\sqrt{2})^{-1} \Big[\sqrt{3} \Big(-t_5^a + \frac{t_5^b + t_5^c} {2} \Big)-\Big(t_2^a - \frac{t_2^b + t_2^c} {2} \Big)\Big] \\ \nonumber
 t_y &=& -\frac{f}{\sqrt{2}} [(t_5^b - t_5^c) + \sqrt{3} (t_2^b - t_2^c) ] \\  \nonumber
 \Delta &=& -(t^a_4 +2t^b_4)- (2)^{-1}[\varepsilon_1+\varepsilon_2- (t^a_6 +2t^b_6)+2\sqrt{3}(t^a_3 +t^b_3+t^c_3)-(t^a_1 +2t^b_1)] \\ \nonumber
  \alpha_1 &=& \frac{1}{2} \Big[\sqrt{3} (t_1^b - t_1^a) -\sqrt{3} (t_6^a - t_6^b) -2 \Big(t_3^a-\frac{t_3^b+t_3^c}{2} \Big) \Big] \\ \nonumber
 \alpha_2 &=& -\sqrt{3} (t_4^a - t_4^b) \\ 
 \delta &=& (\sqrt{2})^{-1} [\sqrt{3}(t^a_2 +t^b_2+t^c_2) - (t^a_5 +t^b_5+t^c_5) ] .
\end{eqnarray}
%
%
 \begin{figure}[t]
\centering
\includegraphics[scale=0.2]{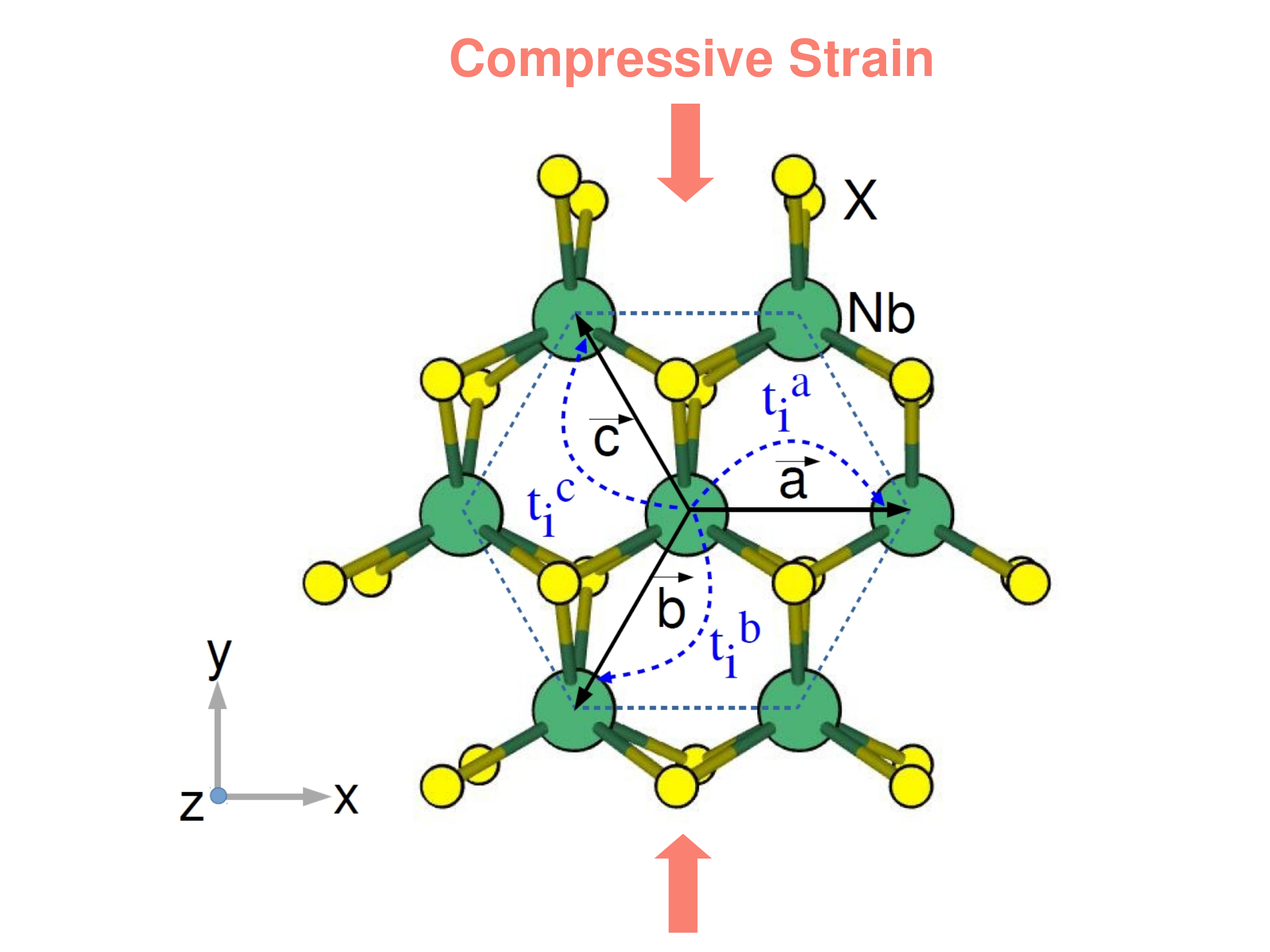}
\includegraphics[scale=0.20]{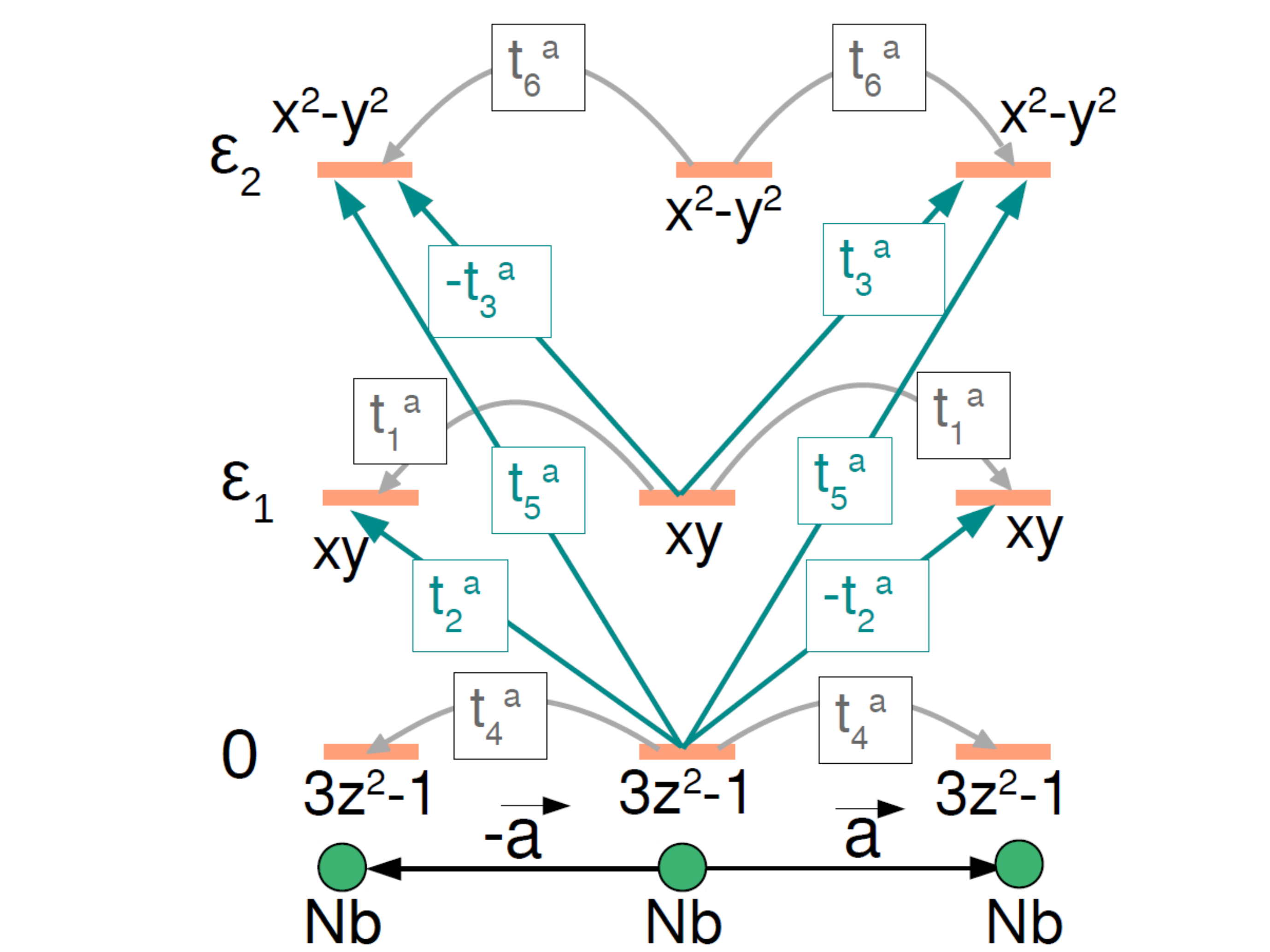}
\includegraphics[scale=0.20]{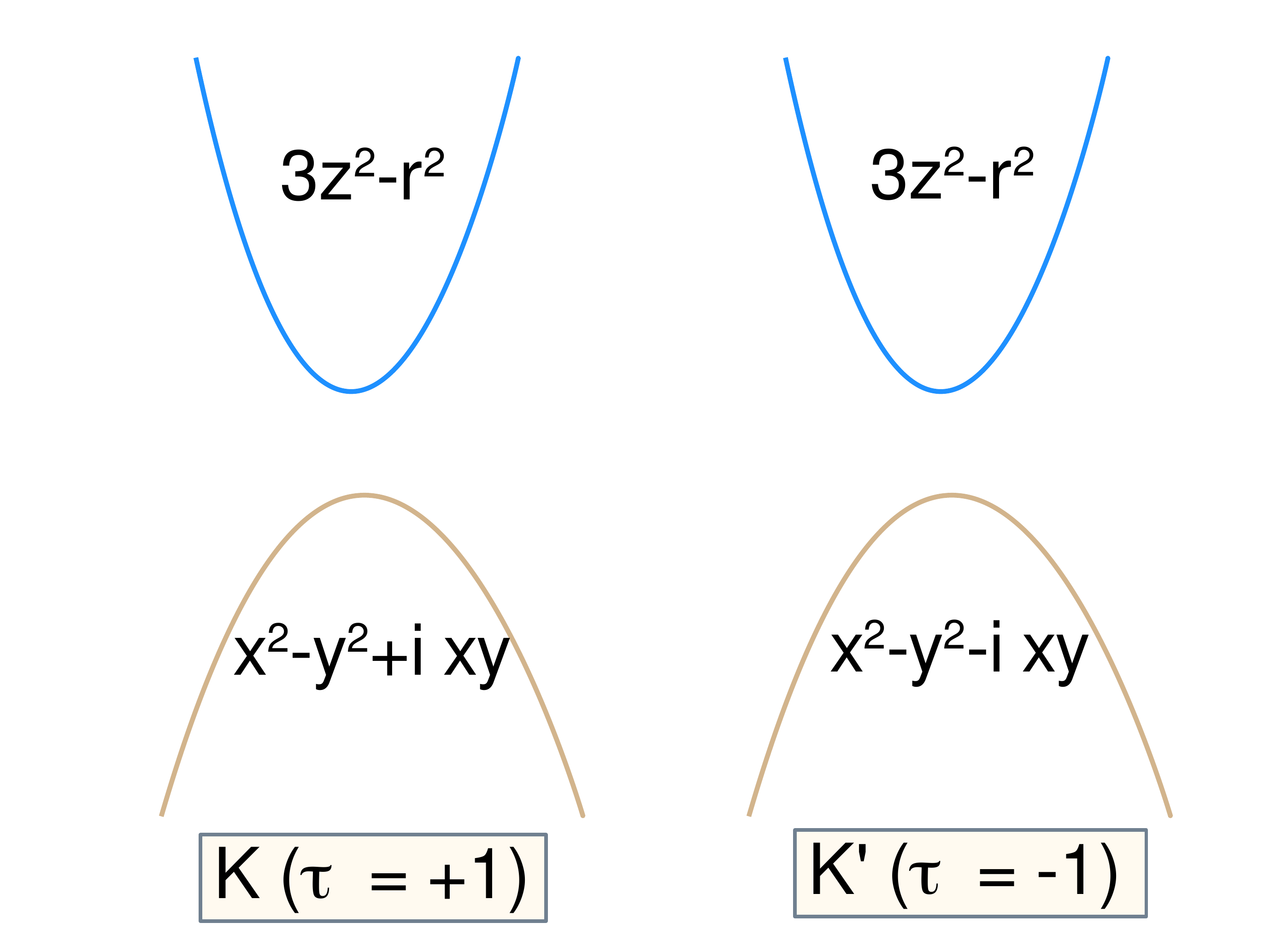} 
\caption {({\it Left}) Tight-binding hopping integrals $t_i^\eta$ between the six neighboring Nb atoms ($i = 1- 6$) along the direction $\vec \eta = \vec a, \vec b, \vec c$. Orange arrows indicate the uniaxial strain direction considered in our work. ({\it Middle}) The Nb-Nb hopping integrals illustrating the left-right asymmetry for hopping along the $\pm \hat a$ direction. ({\it Right}) The valence and the conduction band characters near the valley points in the unstrained TMDCs.
}
\label{fig1}  
\end{figure}
It is interesting to point out that with the help of the above relations, showing the explicit dependences of the various Hamiltonian parameters on the TB hopping, it can be shown that in presence of $C_{3z}$ symmetry of the unstrained structure $\alpha_1$, $\alpha_2$ and $\delta$ become zero and the parameter $t_x$ becomes identical to $t_y$, viz., $t_x = t_y$. This is due to the fact that in presence of $C_{3z}$ symmetry, the hopping parameters $t_i^\eta$ along $\eta= \vec a, \vec b,$ and $\vec c$ are not independent of each other, but they also follow the three-fold rotational symmetry of the structure \cite{ohe_big}, which leads to the desired result.
These parameters, however, become non-zero if the $C_{3z}$ symmetry is broken, as is the case for the strained structure.
These strain induced parameters play the crucial role in driving the OGME, as discussed in the main paper,  emphasizing the importance of the broken $C_{3z}$ symmetry by strain in dictating the OGME in Nb$X_2$.

While in absence of strain, $t_x = t_y =t$ and $\Delta$ describe the electronic structure for the unstrained system, presence of strain leads to $t_x \ne t_y$, and we introduce the parameter $\kappa$ to describe this hopping asymmetry, viz., $t_{x,y} = t \pm \kappa {\cal S}$. Also the last three parameters in Eq. (\ref{para}), viz., $\alpha_1$, $\alpha_2$ and $\delta$, are non-zero in presence of strain, and the leading terms in these three parameters are linear in strain, viz.,
\begin{eqnarray}
 \alpha_1 = \beta_1 {\cal S} , ~~ \alpha_2 = \beta_2 {\cal S}, ~~ \text{and} ~~~
 \delta &=& \gamma {\cal S}. 
\end{eqnarray}
Thus, in summary, the two parameters $t$ and $\Delta$ describe the electronic structure of the unstrained system, while in presence of strain we need the additional parameters $\beta_1, \beta_2, \kappa$, and $\gamma$, thereby resulting in six independent parameters in the model to describe the 
effect of the uniaxial strain. All these six parameters are tabulated in the main paper for monolayer Nb$X_2$, the subject of the present work.

For a general strain condition, the Hamiltonian can be written in terms of 
a strain independent part ${\cal H}_0$ and a strain dependent part $ {\cal H}_S$,
\begin{eqnarray}     \label{Hstrain}
{\cal H} (\vec q ) = {\cal H}_0 +  {\cal H}_S , 
\end{eqnarray} 
where the second term can be expressed in terms of the strain tensor for small strain:
 $ {\cal H}_S = \sum_{ij} \varepsilon_{ij} {\cal H}_1^{ij} $,
where $\varepsilon_{ij}$ is the  $3 \times 3$ strain tensor,  the nine 
${\cal H}_1^{ij} $ matrices are strain-independent, and $i, j = 1, 2, 3$ are the indices for the cartesian coordinates. 

We have considered a special case of the strain
in our work, where there is only a strain component present along the $\hat y$ direction, i.e., 
$\varepsilon_{ij} = {\cal S}\  \delta_{i2} \  \delta_{j2}$. 
The magnitude of the strain is denoted by ${\cal S}$, and
tensile strain by standard convention is defined as positive and compressive strain as negative. 
For this strain case, the Hamiltonian assumes a simple form, which can be written compactly in terms of the pseudo-spin Pauli matrices. From Eqs. (\ref{HK}-\ref{Hstrain}), we obtain straightforwardly the result
\begin{eqnarray} \label{Htot}    
{\cal H} (\vec q ) &=& {\cal H}_0 + S {\cal H}_1 , \\ \nonumber
{\cal H}_0 &=& \vec d \cdot \vec \sigma,  \\  \nonumber
{\cal H}_1 &=& d_0 \sigma_0 + \vec d_1 \cdot \vec \sigma,
\label{HS}
\end{eqnarray} 
where $\sigma_0$ and $\vec \sigma$ are respectively the identity matrix and the Pauli matrices in the pseudo-spin basis $\ket v$ and $\ket c$,
and the coefficients are:  
$\vec d = (\tau t q_x a, -t q_y a, -\Delta/2)$, $ d_0 = \tau q_x a \beta_+/2$, and
$ \vec d_1 = (\tau \kappa q_x a + \gamma, \kappa q_y a, \tau q_x a \beta_-/2)$.
The Hamiltonian (\ref{Htot}) is our desired two-band valley-orbital model, which we studied in the main paper.



\section{Effect of Spin-orbit coupling on GME}

In this section, we explicitly consider the effect of spin-orbit coupling (SOC) within the valley model and show that even in presence of SOC, the spin contribution to the gyrotropic magneto-electric effect (GME) vanishes due to the well known Ising like form of the SOC \cite{Xiao2013} in TMDCs and, therefore, the orbital GME continues to be the dominating effect.

As discussed above, the valence and the conduction bands near the valley points are well described by the wave function characters, $\ket v = (\sqrt{2})^{-1} (|x^2-y^2 \rangle + i \tau |{xy} \rangle)$ and $\ket c = |3z^2-r^2 \rangle$.
Within this subspace, considering also the spin degrees of freedom, viz.,  $\{  \ket{ v \uparrow}, \ket {c \uparrow}, \ket{ v \downarrow}, \ket{ c \downarrow} \} $, the SOC takes the following form
\begin{equation} \label{SOC}
 \mathcal{H}_{\rm SOC} =\frac{\tau \lambda} {2}  (\sigma_z+1) \otimes s_z,
\end{equation}
Here $\vec s$ and $\vec \sigma$ are respectively the Pauli matrices for the electron spin and the orbital pseudo-spins, $\lambda$ is the strength of the SOC, and $\tau$ is the valley index. Note the block diagonal form of $\mathcal{H}_{\rm SOC}$, representing Ising SOC, which does not allow for any spin mixing. We show that this peculiar form of SOC plays an important role in vanishing spin contribution to the GME.    

We first construct the total Hamiltonian by adding the SOC term to the valley-orbital model, given in Eq. (\ref{Htot}). In the basis $\{  \ket{ v \uparrow}, \ket {c \uparrow}, \ket{ v \downarrow}, \ket{ c \downarrow} \} $, the total Hamiltonian has the following form
\begin{equation} \label{withSOC}
\mathcal{H}_{\rm tot} =  \mathcal{H} (\vec q) \otimes I_s + \frac{\tau \lambda} {2}  (\sigma_z+1) \otimes s_z.
\end{equation}
Here $I_s$  is the $2 \times 2$ identity operator in the electron spin space. The SOC term splits the spin-degenerate valence bands into fully polarized $S_z = \uparrow$ and $\downarrow$ bands, as shown schematically in Fig. \ref{fig2} (a) and at the inset of Fig. \ref{fig2} (b), where the expectation value of the $S_z$ operator for the bottom valence band at the $K$ and $K'$ valleys is shown along the $q_x$ direction. As evident from this figure, the valence bands have pure $S_z$ character, and there is no spin mixing between the valence bands.

We have numerically computed the spin and orbital contributions to the GME. 
The results of our calculations are shown in Fig. \ref{fig2} (b) as a function of (VBM-$E_F$)$/\Delta$, where $E_F$, VBM and $\Delta$ are respectively the Fermi energy, the valence band maximum, and the energy gap at the valley points. As clear from this figure, the spin contribution to the GME vanishes for all values of (VBM-$E_F$)$/\Delta$, which can also be inferred from the pure $S_z$ characters of the valence bands, indicating no momentum space variation of $S_z$ near the valley points (see inset of Fig. \ref{fig2} (b)) in contrast to the orbital moment, as discussed in the main text. Note that, the negligibly small spin contributions in the real systems may be attributed to the higher order effects, which are ignored in this model, that may lead to small spin mixing between the valence bands, giving rise to a negligibly small spin contribution.  
 In essence, the spin moments contribute zero to the GME 
within the Ising-like model Eq. (\ref{SOC}), 
because the 
spin moments in the Brillouin zone have a fixed value for each band ($S_z = \pm 1$ as indicated in the inset of Fig. \ref{fig2}). As a result, the net spin moment is unaltered by shifting of the Fermi surface due to the applied electric field.

Unlike the spin contribution, the orbital contribution remains significant and, therefore, dominantes the GME. As expected, the orbital contribution is zero when the Fermi energy is at the top of the valence band [(VBM-$E_F$)$/\Delta$ = 0 in Fig \ref{fig2} (b)], representing the insulating case and, then, starts to increase as the Fermi energy gets deeper into the valence bands [(VBM-$E_F$)$/\Delta > 0$]. The vertical dashed line in Fig. \ref{fig2} (b), corresponds  to the typical hole concentration ($\sim 0.25 $) per each valley, for which the magnitude of the orbital contribution is about 90 $\mu_B$ V$^{-1}$ \AA$ ^{-1}$. We note that this magnitude is although smaller than the case without SOC, still it is significantly large and, more importantly, the dominant contribution to the GME.

 \begin{figure}[t]
\centering
\includegraphics[scale=0.4]{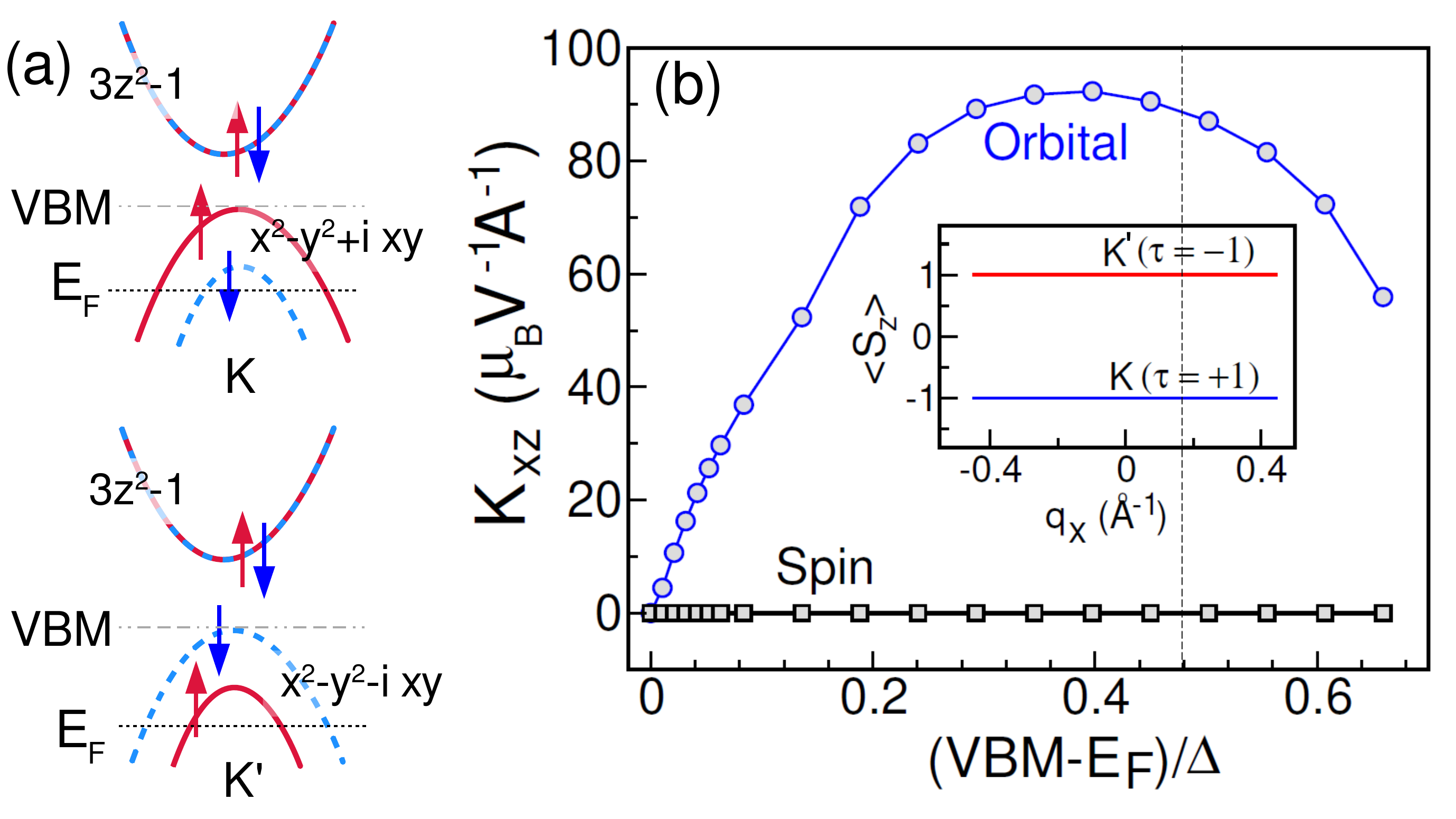}
\caption { (a) Schematic illustration of the spin split valence bands around the valley points. The valence band maximum (VBM) and the Fermi energy $E_F$ are indicated as light and dark dashed lines. (b) The variation of the orbital and the spin contributions to the GME in presence of SOC as function of (VBM-$E_F$)$/\Delta$. The vanishing spin contribution can be inferred from the pure spin character of the valence band, as shown at the inset of (b). The inset shows the expectation value of $S_z$ operator in units of $(\hbar/2)$.
 The dominant orbital contribution in presence of SOC is clear from this plot. Here, SOC $\lambda$ is taken to be $0.08$ eV, relevant for NbSe$_2$.}
\label{fig2}  
\end{figure}


\section{Additional Density functional results}


\subsection{DFT Methods}
All density-functional  calculations in the present work are performed with the relaxed structures of monolayer Nb$X_2$, $X =$ S, Se. 
A uniaxial strain along $\hat y$ direction is applied for each case and all atomic positions are relaxed, keeping the unit cell parameters fixed corresponding to the applied strain,  
 until the Hellman-Feynman forces on each atom becomes less than 0.01 eV/\AA  ~using Vienna {\it ab initio} simulation package (VASP) \cite{vasp}.

With the relaxed structures of monolayer Nb$X_2$, $k$-space orbital moment and the GME are calculated using QUANTUM ESPRESSO and Wannier90 codes \cite{QE, w90_code}, both in presence and absence of strain. 
Self-consistency is achieved using fully relativistic norm-conserving pseudopotentials for all the atoms with a convergence threshold of 10$^{-7}$ Ry. 
The {\it ab-initio} wave functions are projected to maximally localized
Wannier functions \cite{MLWF} using the Wannier90 code \cite{w90_code}.
In the disentanglement process, as initial projections, we have chosen 26 Wannier functions per unit cell which include the $d$ orbitals of Nb and $s$ and $p$ orbitals of $X$ atoms, excluding the rest.
 After the disentanglement is achieved, the wannierisation process is converged to 10$^{-10}$ \AA$^2$ and, then, the $k$-space orbital moment $\vec M (\vec k)$ and the gyrotropic response $\tilde K_{xz}$ are calculated. In the
GME calculations, satisfactory convergence was obtained by adopting a  $500 \times 500 \times 1$ $k$-mesh grid size.

\subsection{Bandstructures and Energy contours}

 \begin{figure}[t]
\centering
\includegraphics[scale=0.4]{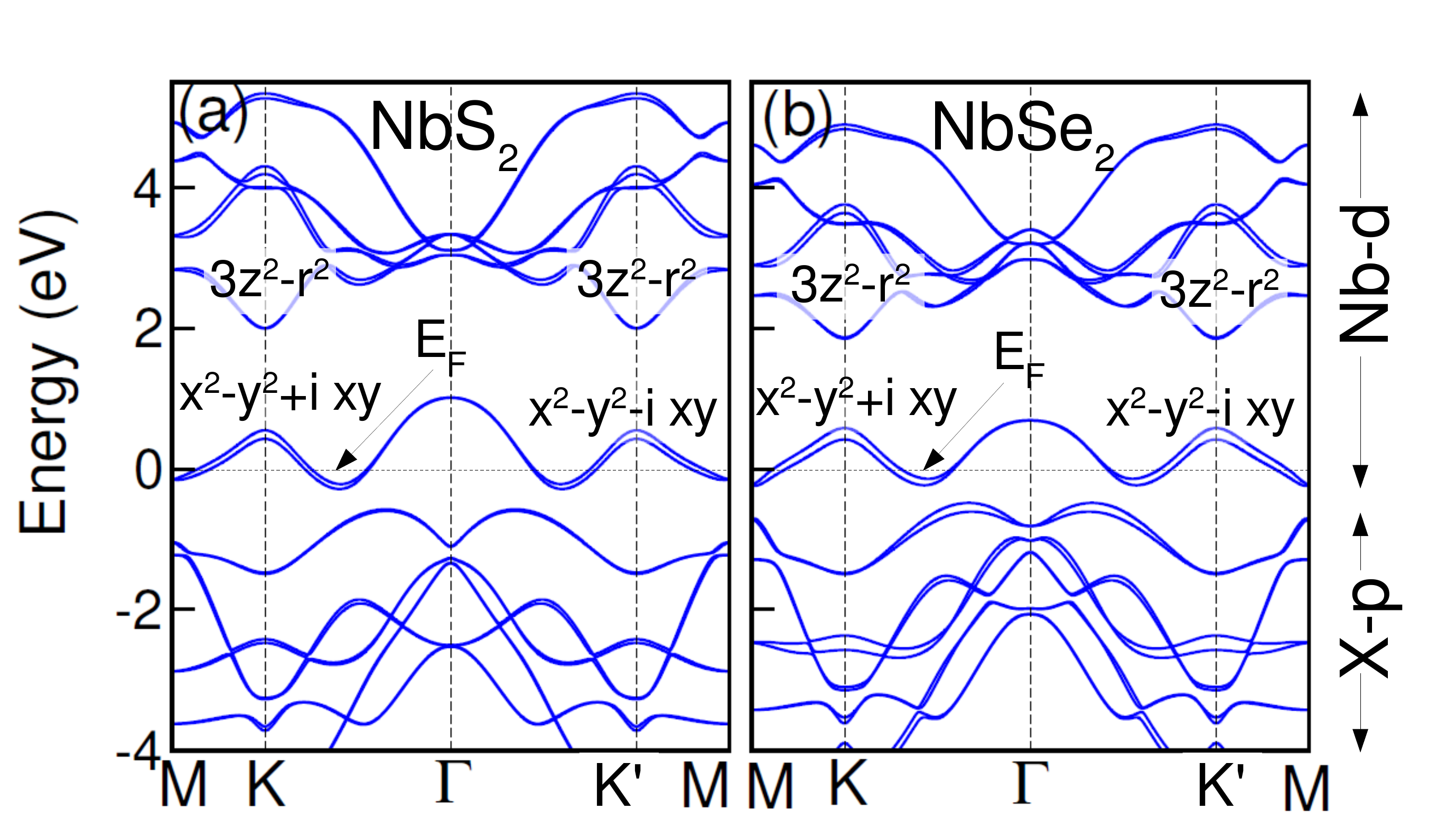}
\caption {Band structures for the unstrained monolayer (a) NbS$_2$ and (b) NbSe$_2$. The orbital characters are indicated in the figure.}
\label{band_wostrain}  
\end{figure}

The band structures of the unstrained monolayer Nb$X_2$, $X =$ S, Se are shown in Fig. \ref{band_wostrain}. As seen from this figure, the valence and the conduction bands near the valley points are predominantly formed by the orbitals $\ket v = (\sqrt{2})^{-1} (|x^2-y^2 \rangle + i \tau |{xy} \rangle)$ and $\ket c = |3z^2-r^2 \rangle$ respectively, where $\tau = \pm 1$ indicate K and K' valleys respectively. The complex $d$ orbitals at the valley points lead to the robust orbital moments, as discussed in the main paper.

In order to understand the changes in the band structure due to the application of a uniaxial strain, the DFT band structures both in presence and absence of strain are compared in Fig. \ref{band} for the monolayer NbSe$_2$. 
The valence band in NbSe$_2$ is half-filled due to the $d^1$ electronic configuration of the Nb atom, leading to one hole in the valence band. 
We estimate the number of holes around the valley points, which is about $n_h \approx 0.25$ for each valley.
As seen from these band structures, the positions of the valence band maxima at $\Gamma$ and $K / K'$ shift with strain. This results in a change in the number of holes around the valley points for the two different strain types. The computed number of holes for tensile and compressive cases are respectively $n_h \approx  0.2$ and $ 0.3 $ for each valley. To keep our model simple, we did not include this effect in our valley-orbital model. As discussed in the main paper, this may be a contributing factor to the asymmetry in the OGME magnitude for compressive and tensile strain.

 \begin{figure}[t]
\centering
\includegraphics[scale=0.5]{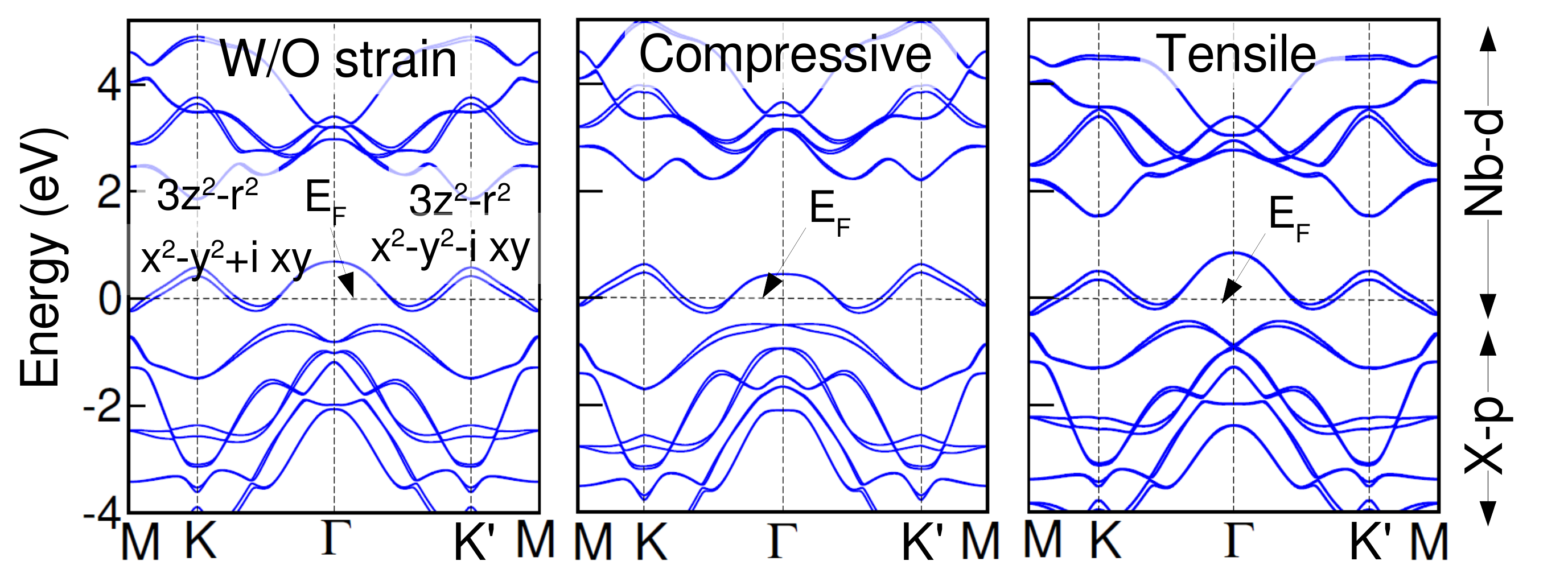}
\caption {Comaprison of the DFT band structures in absence and presence of strain. The band structures of NbSe$_2$ is shown ({\it left}) without the strain, in presence of compressive ({\it middle}) and tensile ({\it right}) strain. $\pm 5 \%$ strains are used for the calculations. As apparent from the band structures, the number of holes at the valley points differ for the two different strain types.}
\label{band}  
\end{figure}

 \begin{figure}[h!]
\centering
\includegraphics[scale=0.5]{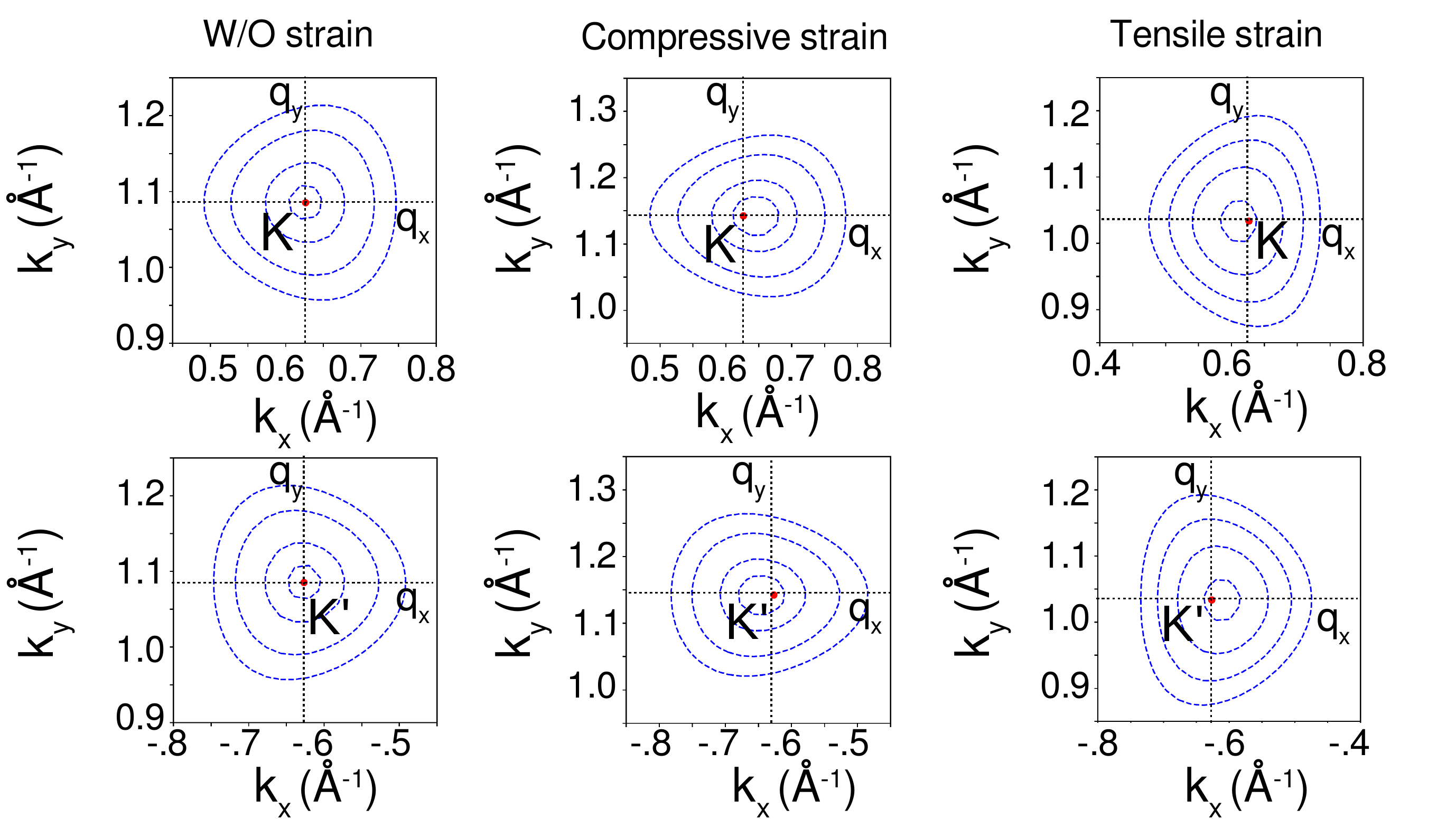}
\caption {Energy contours near the valley points in NbS$_2$ in absence and presence of strain. The circular energy contours, centered around the valley point, in the absence of strain, become elongated along $\hat x$ or $\hat y$ in presence of the compressive and tensile strain respectively. Also, the center of the ellipse is shifted from the valley points along $\hat x$, which plays the crucial role in driving the OGME. 
}
\label{contour}  
\end{figure}

The strain induced modifications in the band structure can be, further,  understood from the the constant 
energy contours in the neighborhood of the the valley points, as shown in Fig. \ref{contour} both in absence and presence of strain. As shown in the figure, in absence of strain the energy contours are circular in shape and are centered at the valley point. In presence of strain, the circular contours distort in shape and become elliptical. More interestingly, the center of the ellipse is also shifted away from the valley point. The shift of the center of the ellipse is opposite for two different strain types as well as for the two valleys for a fixed type of strain. 
The shift of the center of the ellipse plays a crucial role in both OGME and its switching, as discussed in the main paper.



\end{widetext}

\end{document}